\documentclass[epsfig,preprint,showpacs]{revtex4} 
\input{epsf.tex}
\usepackage{amssymb,epsfig}

\newcommand{\One}{1\kern-4.5pt1}
\newcommand{\lapprox}{\raisebox{-0.5ex}{$\ 
\stackrel{\textstyle<}{\textstyle\sim}\ $}}

\begin{document}
\renewcommand{\topfraction}{.95}
\renewcommand{\textfraction}{.1}

\title{Interplay of size and Landau quantizations in the de Haas-van Alphen oscillations of metallic nanowires}
\author{A. S. Alexandrov$^1$,  V. V. Kabanov$^{2}$, and I. O. Thomas$^1$}

\affiliation{$^1$Department of Physics, Loughborough University,
Loughborough, United Kingdom\\ $^{2}$Josef Stefan Institute 1001,
Ljubljana, Slovenia}

\begin{abstract}
 We examine the interplay between size quantization and Landau quantization in the De Haas-Van Alphen oscillations of clean, metallic nanowires in a longitudinal magnetic field for `hard' boundary conditions, i.e. those of an infinite round well, as opposed to the `soft' parabolically confined boundary conditions previously treated in Alexandrov and  Kabanov (Phys. Rev. Lett. {\bf 95}, 076601 (2005) (AK)).  We find that there exist {\em two} fundamental frequencies as opposed to the one found in bulk systems and the three frequencies found by AK with soft boundary counditions.  In addition, we find that the additional `magic resonances' of AK may be also observed in the infinite well case, though they are now damped.  We also compare the numerically generated energy spectrum of the infinite well potential with that of our analytic approximation, and compare calculations of the oscillatory portions of the thermodynamic quantities for both models.
\end{abstract}

\pacs{75.75.+a,
73.63.Nm,
73.63.b}

\maketitle
\section{Introduction}

Since their discovery in 1930 \cite{DeHaas}, and subsequent explanation by Onsager \cite{Onsager} and  Lifshitz and Kosevitch \cite{Lifshitz} in terms of the interplay between the  Landau quantization and the Fermi surface, the de Haas-Van Alphen (DHVA) oscillations of the thermodynamic potential, magnetization and related quantities has acquired a long history of use as a method of probing the structure of the Fermi surface of metals at low temperatures and large magnetic fields. (A summary of the first thirty-eight years of study in this field may be found in Gold \cite{Gold}. More contemporary reviews of the area may be found in Shoenberg \cite{Shoenberg}, Singleton \cite{Singleton} and Kartsovnik \cite{Kartsovnik}, with the latter two focusing on applications to quasi-2D organic compounds.)

The question of boundary conditions and how confinement affects the behavior of electrons in longitudinal magnetic fields (and hence the DHVA effect) has long been a question of interest to the condensed matter community (see, for example, Fock \cite{Fock}, Landau \cite{Landau}, Darwin \cite{Darwin} and also the discussion in Dingle's classic paper \cite{Dingle1} on the effects of boundary conditions on the density of states in large systems).  This issue is of obvious importance in the study of nanowires, where the system of interest is confined within a narrow, cylindrical potential, which  in this case is difficult to treat analytically owing to the  difficulties in approximating the eigenvalues of the Schr\"odinger equation given by the zeroes of the confluent hypergeometric function.  However, in the weak field limit, the eigenvalues are well approximated by the zeroes  of the Bessel function, and so Dingle \cite{Dingle2} was able to calculate the effect of such a confinement in a weak magnetic field, through treating the latter as a small perturbation to the energy levels and showing that at low fields, one should expect to observe oscillations that are primarily due to the passage of the size quantized energy levels above the Fermi energy. 
  Bogachek and Gogazde \cite{Bogacheck} calculated the contribution to the thermodynamic potential of those electrons near to the surface of the wire that are on a `grazing' trajectory with respect to the surface, and argued that this is in fact a more dominant contribution to the thermodynamic potential than the Dingle-type oscillations, as the latter are suppressed by specular reflection under normal circumstances.  The contribution of these surface levels corresponds to that induced by the Aharonov-Bohm effect.
  This effect was observed, along with others ascribed to Dingle-type oscillations by Brandt {\em et al.} \cite{Brandt} in a study of the magnetoresistance of bismuth wires of micrometer thickness.  Further confirmations of the presence of the low field Aharonov-Bohm effect before the onset of Shubnikov-De Haas oscillations in the magnetoresistance  comes from the study of progressively narrower Bi nanowires \cite{Huber}, as well as in carbon nanotubes \cite{Fujiwara,Roche}.

It is likely that at higher fields for wires of very narrow radius, we might expect additional effects of the interplay between size quantization and Landau quantization to become apparent.  There is a need for theoretical treatments of this region, and in what follows we make one of the first attempts at doing so outside of an idealized 1-D limit, such as that used in Sineavsky {\it et al.} \cite{Sineavsky}. 


Alexandrov and Kabanov (AK)\cite{Alexandrov1} have recently examined the DHVA oscillations present in a system of electrons confined within a parabolic confining potential (the Fock-Darwin or confinement model \cite{Fock,Darwin}). They have found that there  exist  {\em three} characteristic frequencies of DHVA oscillation, and  also that the thermodynamic potential displays `magic resonances' at certain values of the field where the amplitude is greatly enhanced.  In a brief note at the end of this Introduction, we demonstrate that despite being due to a divergent cotangent function in our expression for the oscillatory thermodynamic potential $\tilde \Omega$, the behavior of the thermodynamic potential is still analytically well defined -- this is important if the AK result is to be properly understood.

Our paper focuses on a system of free electrons confined within a cylindrical infinite well potential and subject to a longitudinal magnetic field.  Using a well known asymptotic description of the wavefunction of such a system, we derive a non-linear energy spectrum and confirm it by deriving a similar result through a semiclassical approximation.  By linearizing these spectra near the Fermi energy, we find that we acquire expressions for the energy similar to that of the Fock-Darwin model, and so we may motivate an improvement to the asymptotic and semi-classical models that brings them closer to the energy spectrum at larger values of the field than was previously possible.  The oscillations in  the thermodynamic potential are calculated, and we find not only that the additional characteristic frequencies of the system are modified, but that the position of the magic resonances has altered and that they are somewhat damped.  Numerical plots of the thermodynamic quantities calculated are given for various values of the field.

Finally, we examine how well the modified asymptotic approximates the exact spectrum of the problem.  We find that it is a reasonable approximation for low values of the magnetic quantum number $m$ at large $\omega_c={eB}/{m^*c}$; however, it is inaccurate for large values of $m$, reflecting orbits that are closer to the edge of the wire.  This is likely because the asymptotic that we are using is not valid in that limit (similarly, the semiclassical approximation breaks down here since it is too close to a turning point of the classical electron trajectories).  However, the actual boundary conditions of the system are most likely intermediate between those of the infinite cylindrical well and those of the Fock-Darwin confinement model, and so it is perhaps reasonable to suspect that the breakdown of the approximation, which is likely due to the `hardness' of the boundary conditions in the infinite well case, should not unduly restrict the application of the nonlinear spectrum to real systems.

In what follows, we assume that the system of interest is a long, clean metallic nanowire in a longitudinal magnetic field ${\bf B}$ that is parallel to the direction of the wire (taken to be the $z$ axis of a cylindrical coordinate system).  We take the electron mean free path, $l=v_F \tau$ to be comparable to or larger than the radius of the wire $R$, but not the length $L$.  We also assume that the electron wavelength near the Fermi level is very small in these metallic nanowires, such that $L\gg l \lapprox R \gg 2\pi\hbar/(m^* v_F)$; here and previously, $v_F$ is the Fermi velocity and $m^*$ is the band mass in the bulk metal.

From henceforth, we also take $\hbar=c=m^*=L=1$.

\subsection{On the divergence of the `magic resonances' in the AK result}

The AK \cite{Alexandrov1} result describes the DHVA oscillations of a three-dimensional electron exposed to a magnetic field within a two-dimensional parabolic confining potential \cite{Fock,Darwin}.  This system is an isotropic planar harmonic oscillator whose frequency in the absence of a magnetic field is taken to be $\omega_0$.  Due to the effects of confinement, the DHVA oscillations of the system aquire three characteristic frquencies of oscillation: $2\omega$, $\omega^+$ and $\omega^-$, where $\omega^2=
\omega_0^2+\omega_c^2/4$ and $\omega^\pm=\omega\pm \omega_c/2$. The amplitudes are given by 
\begin {equation}
A_r(x,y)= {T (2x)^{1/2} \over{2\pi r^{3/2} \sinh[\pi^2Tr/( x)]}}
\cot \left({\pi r y\over{2x}}\right) \label{eqn:confamp},
\end{equation}

\noindent where $x=\omega$ and $y=\omega^\pm$ for the the amplitudes of the oscillations with frequency $2\omega$, and $x=\omega^\pm/2$ and $y=2\omega$ for those with frequency $\omega^\pm$.  As has been previously noted, these amplitudes (\ref{eqn:confamp}) contain a cotangent function which gives rise to `magic resonances' for certain values of the field where  the condition $2\omega/(\omega^\pm) =(q+2)/r$ is satisfied. If this is the case then the cotangent becomes infinite provided that $q$ is an integer.

However, despite what one might naively assume, this apparent divergence is in fact analytically well defined.  Let us examine the following harmonics: $r=1,\, q=1$ and $r=3,\,q=9$.  From our condition for the divergence of the cotangent, we have $\omega=\frac{3}{2}\omega_c$ and hence $\omega^{-}=\frac{2}{3}\omega$. Let us set $\omega=1$, and then set $\omega^{-}=\frac{2}{3}(1-\delta)$, where $\delta\rightarrow+0$. 

For our $r=1$ harmonic, we shall examine the $A_1(\frac{\omega^-}{2},2\omega)$ amplitude, and for our $r=3$ harmonic, we shall examine the $A_3(\omega,\omega^-)$ amplitude.  We acquire:
\begin{eqnarray}
A_{1}\left({\frac{\omega^-}{2}},2\omega\right)&=&A_1\left(\frac{1}{3}(1-\delta),2\right) \nonumber \\
&=& { T (2/3)^{1/2} \over{2\pi  \sinh[3\pi^2T]}}
\cot \left({3\pi \over{1-\delta}}\right) \label{eqn:A1mr}
\end{eqnarray}

\noindent and
\begin{eqnarray}
A_{3}\left({\omega},\omega^-\right)&=&A_3\left(1,\frac{2}{3}(1-\delta)\right) \nonumber \\
&=& {T (2/3)^{1/2} \over{6\pi \sinh[3\pi^2T]}}
\cot \left(\pi(1-\delta) \right) .\label{eqn:A3mr}
\end{eqnarray}

\noindent Expanding out the cotangent from (\ref{eqn:A1mr}), we find
$\cot \left({3\pi}/({1 -\delta})\right)\approx {1}/({3\pi\delta})$ while doing the same for (\ref{eqn:A3mr}) gives $\cot\left(\pi\left(1-\delta\right)\right) \approx-{1}({\pi\delta})$.  Substituting these values back into (\ref{eqn:A1mr}) and (\ref{eqn:A3mr}) respectively, we find that:
\begin{equation}
A_1\left({\frac{1}{3}(1-\delta)},2\right)
=-A_3\left(1,\frac{2}{3}(1-\delta)\right)
={T (2/3)^{1/2} \over{6  \pi^2   \sinh[3\pi^2T]}}\frac{1}{\delta}\equiv \frac{A}{\delta}.
\end{equation}

We may now write, where $\tilde\Omega_{res}$ is the oscilliatory component of the AK result near the magic resonance:
\begin{eqnarray}
\tilde\Omega_{res}&=&A_{3}\left({\omega},\omega^-\right)\sin\left(\frac{3\mu\pi}{\omega} -\frac{3\pi(\omega^+ + \omega^-)}{2\omega} -\frac{\pi}{4}\right)
+A_{1}\left({\frac{\omega^-}{2}},2\omega\right)\sin\left(\frac{2\mu\pi}{\omega^-} -\frac{3\pi(\omega^+ + \omega^-)}{\omega^-} -\frac{\pi}{4}\right)\nonumber\\
&=&-\frac{A}{\delta}\sin\left(3\mu\pi-
\frac{3\pi}{2}\left(\omega+\frac{2}{3}(1-\delta)\right)-\frac{\pi}{4}\right)
+\frac{A}{\delta}\sin\left(\frac{3\mu\pi}{1-\delta}-
\frac{3\pi}{2(1-\delta)}\left(\omega+\frac{2}{3}(1-\delta)\right)-\frac{\pi}{4}\right)\nonumber.
\end{eqnarray}

\noindent Defining $C\equiv 3\mu\pi-\frac{3}{2}\pi(\omega^+ + \frac{2}{3})$ and expanding out the terms in the denominators, we arrive at:
\begin{eqnarray}
\tilde\Omega_{res}&=&AC\cos\left(C-\frac{\pi}{4}\right).
\end{eqnarray}

\noindent Note that the singular terms are no longer present, so that the thermodynamic potential in the region of the resonance is now finite.  However, it has also been enhanced by a factor of $C$, which is of the order of the Fermi energy $\mu\gg1$, and so remains rather large.  We would expect the same analysis to hold true for the other magic resonances.


\section{Cylindrical confinement and the DHVA effect}

The Schr\"odinger equation in polar coordinates for an electron moving in a longitudinal magnetic field $B$ parallel to the $z$ axis is:
\begin{equation}
-\frac{1}{2} \left[\frac{1}{r} \frac{\partial}{\partial r}\left(r\frac{\partial\psi}{\partial r}\right) + \frac{\partial^2 \psi}{\partial z^2} + \frac{1}{r^2}\frac{\partial^2 \psi}{\partial \phi^2} \right] + \frac{i\omega_c}{2} \frac{\partial \psi}{\partial \phi} + \frac{\omega_c^2 r}{8}\psi=E\psi.\label{eqn:3dSchr}
\end{equation}

\noindent Following for example (though note the difference in convention regarding $e$)  Landau and Lifshitz \cite{Landau&L} we seek a solution in the form 
\begin{equation}
\psi=\frac{1}{\sqrt{2\pi}}R(r)e^{im\phi}e^{ip_z},
\end{equation}

\noindent where $R(r)$, after redefinition in terms of $\xi=(\omega_c/2)r^2$ is the radial function:
\begin{equation}
R(\xi)=e^{-\frac{\xi}{2}}\xi^{\frac{|m|}{2}}M(-(\beta-\frac{|m|}{2}-\frac{1}{2}),\,|m|+1,\xi),
\end{equation}

\noindent with $\beta=\omega_c^{-1}[E-(p_z^2/2)] + m/2$.  $M(-(\beta-\frac{|m|}{2}-\frac{1}{2}),\,|m|+1,\, \xi)$ is a confluent hypergeometric function.  The zeroes of this are the eigenvalues of the Schr\"odinger equation for cylindrically confined electrons.  If we write $\beta-\frac{|m|}{2}-\frac{1}{2}$ as $-a$, following the notation of Abramowitz and Stegun \cite{Abramowitz}, the energy levels of the particle are given by \cite{Landau&L}
\begin{equation}
E=\frac{k^2}{2}+\omega_c \left(-a + \frac{|m|-m+1}{2}\right).\label{eqn:xctel}
\end{equation}

If the wave-function is finite everywhere, then $-a$ is an integer, and the eigenfunctions of the equation are the Laguerre polynomials.  However, if this is not the case, and it becomes zero at some finite radius $R$, then $-a$ is a positive, real number.  In order to proceed analytically in these cases, particularly where $R$ is small, we must therefore use various approximation or numerical methods so as to obtain the energy spectrum.


\subsection{Asymptotic approximation}
           
For large enough values of $a$, where $a$ is negative, the confluent hypergeometric function $M(a,b,\xi)$ may be approximated by the following asymptotic form \cite{Abramowitz}:
\begin{equation}
M(a,b,\xi)=\Gamma(b)e^{{\xi}/{2}}\left(\frac{1}{2}b\xi-a\xi\right)^{(1-b)/2}\pi^{-\frac{1}{2}}\cos\left(\sqrt{2b\xi-4a\xi}-\frac{1}{2}b\pi+\frac{1}{4}\pi\right) 
\label{eqn:cosapp}
\end{equation}

\noindent where (as before) we take $a= -(\omega_c^{-1}[E-\frac{p_z^2}{2}] + \frac{m}{2}-\frac{|m|}{2}-\frac{1}{2})$, $b=|m|+1$ and $\xi={\omega_cR^2}/{2}= ({\pi}/{2\omega_s})^2\mu\omega_c $, with $R$ being the radius of the nanowire and $\omega_s=\pi v_F/2R$.  In so choosing the value of $\xi$, we are imposing `hard' boundary conditions by stating that the wavefunction {\it must} be zero at $R$.

Our approximation has zeroes when the following is true:
\begin{equation}
\sqrt{2b\xi-4a\xi}-\frac{1}{2}b\pi+\frac{1}{4}\pi=\frac{2n+1}{2}\pi,
\end{equation}

\noindent and so, after some algebra, we obtain the following approximate expression for the eigenvalues, valid wherever the asymptotic (\ref{eqn:cosapp}) is valid:
\begin{equation}
E_{\alpha}=\frac{\omega_s^2}{\mu}\left(n+\frac{|m|}{2} +\frac{3}{4}\right)^2-\frac{\omega_c m}{2}. \label{eqn:asympen}
\end{equation}

 A similar energy spectrum may be derived from the Bohr-Sommerfeld quantization conditions:
\begin{equation}
E=\frac{\omega_s^2}{\mu}\left(n+\frac{\tilde{m}+1}{2}\right)^2 -\frac{\omega_c m}{2}, \label{eqn:semiclassen}
\end{equation}

\noindent where $\tilde{m}=\sqrt{m^2-1/4}$. 

Up to a phase factor this is very similar to (\ref{eqn:asympen}), which is derived from the asymptotic form of the confluent hypergeometric function. In what follows we shall tend to use the spectrum (\ref{eqn:asympen}) unless otherwise stated, since there is a possibility that the Bohr-Sommerfeld conditions do not work as well as they might, as has been noted in the case of a nanowire in a transverse electric field (as observed in \cite{Chaudhuri}, for example).


\subsection{Connecting the approximations with the parabolic confinement model}\label{sec:connect}

We now turn to address what, if anything, (\ref{eqn:asympen}) and (\ref{eqn:semiclassen}) have to do with the confinement potential spectrum (\cite{Alexandrov1,Fock,Darwin} -- here we take $\omega_0$ to be the frequency of the confining potential used in those sources).  For simplicity, we shall set $\omega_c=0$ (i.e. work in the limit of zero magnetic field).  Defining $\xi=E-\mu$, and working where $E\approx\mu$, we may write for our non-linear spectra (\ref{eqn:asympen},\ref{eqn:semiclassen}):
\begin{equation}
\xi=\frac{\omega_s^2}{\mu}{\mathcal A}^2 -\mu \approx 2\omega_s{\mathcal A}-2\mu \label{eqn:linearised}
\end{equation}

\noindent where 
\begin{equation}
{\mathcal A}=\left\{\begin{array}{cc}
               \left(n+\frac{|m|}{2} +\frac{3}{4}\right) & \mbox{asymptotic}\\
	       &\\
	       \left(n+\frac{\tilde{m}+1}{2}\right) & \mbox{semiclassical}
	       \end{array}\right..
\end{equation}

When $\omega_c=0$, $E\approx\mu$, the spectrum of the parabolic confinement model becomes:
\begin{equation}
\xi=E-\mu=2 \omega_0\left(n+ {|m|+1\over{2}}\right)-\mu.
\end{equation}

So, in this limit, the confinement model, if we take $\omega_0\equiv\omega_s$, and the non-linear approximations are good matches up to an unimportant phase and a factor of $2$ in front of the $\mu$ term.  Since the latter only shifts the value of the energy by a constant, it seems that the results of \cite{Alexandrov1}  are a good approximation of the behavior of the non-linear approximations (\ref{eqn:asympen},\ref{eqn:semiclassen}), very near the Fermi energy, and at zero field.

This motivates a replacement of  $\omega_s$ in (\ref{eqn:linearised}), and hence in (\ref{eqn:asympen}) and (\ref{eqn:semiclassen}), with $\omega$ so as to better approximate the behavior of the system at larger values of $\omega_c$ with these non-linear spectra.  That this is not a bad approximation (at least for small and intermediate values of $m$) is suggested by the numerical analysis of Section \ref{sec:errors}. 

Having made this replacement and restoring the momentum component in the $z$ direction, we may write the spectra of the non-linear model (\ref{eqn:asympen},\ref{eqn:semiclassen}) of electrons confined in a nanowire as:
\begin{equation}
E=\frac{k^2}{2}+\frac{\omega^2}{\mu}{\mathcal A}^2-\frac{\omega_c m}{2}.\label{eqn:nlmod}
\end{equation} 

It is also instructive to compare this with the spectrum used by Bogachek and Gogadze \cite{Bogacheck} and Dingle \cite{Dingle2}.  Perturbation theory for small values of the field gives us the following expression for the energy:
\begin{equation}
E=\frac{\epsilon_{nm}}{2m^*R^2} + \frac{k^2}{2m^*},
\end{equation}

\noindent where in this case we restore $m^*$ for clarity, and 
\begin{equation}
\epsilon_{nm}=\gamma^2_{nm} - 2\phi m +\frac{1}{3}\phi^2\left[1+\frac{2(m^2-1)}{\gamma^2_{nm}}\right],
\end{equation}

\noindent with $\gamma_{nm}$ being the $n$-th zero of the $m$-th order Bessel function, and $\phi={\Phi}/{\Phi_0}$, where $\Phi=\pi R^2B$ and $\Phi_0=hc/e$ (in ordinary units).  Defining a frequency $\nu_s={v_F}/{R}$ our overall expression for the energy becomes:
\begin{equation}
E=\frac{k^2}{2}+\frac{\nu_s^2}{4\mu}\gamma^2_{nm}-\frac{\omega_c}{2} + \frac{\phi\omega_c}{12}\left[1+\frac{2(m^2-1)}{\gamma^2_{nm}}\right].
\end{equation}

Discarding the last term, and noting that the zeros of the Bessel function in our region of interest (as opposed to that of Bogacheck and Gogazde \cite{Bogacheck} and Dingle \cite{Dingle2}) are given by $\pi(n +\frac{|m|}{2} +\frac{3}{4})$ we arrive at our asymptotic expression for the energy at weak field.  It is possible to recover the Aharanov-Bohm oscillations predicted by Bogacheck and Gogazde from the results of our calculations in the next section by examining the limit of small $\omega_c$. However, the oscillations they obtain have the form $\cos(2\pi r\phi)$, whereas ours have the form $\cos(4r\phi/\pi)$.  It seems probable that for the most part this discrepancy due to our use of a different asymptotic approximation of the Bessel function zeroes -- they use an approximation valid for large $m$, and we use one valid for large $n$.


\subsection{The De Haas--Van Alphen effect for the non-linear model}\label{sec:nlDHVA}

We now calculate the oscillatory portion of the thermodynamic potential for the asymptotic version of (\ref{eqn:nlmod}) -- henceforth referred to as the non-linear model -- neglecting the phase term in the quadratic portion since it is much smaller than $n+\frac{|m|}{2}$, so that we have:
\begin{equation}
E_{nmk}=\frac{k^2}{2}- \frac{m \omega_c}{2} + {\omega^2\over{\mu}}(n+{\textstyle\frac{|m|}{2}})^2. \label{eqn:calcasymp}
\end{equation}

\noindent  Applying twice the Poisson's formula to the expression for the thermodynamic potential
\begin{equation} 
\Omega =-T \sum_{\alpha} \ln
[1+e^{(\mu-E_{\alpha})/T}], \label{eqn:thrmpot}
\end{equation}

\noindent
 and
replacing negative $m$ with $-(m+1)$ one obtains
 \begin{equation}
\Omega =\sum
_{r,r'=-\infty}^{\infty}[\Omega^{+}_{rr'}+\Omega^{-}_{rr'}],\label{eqn:omegapm}
\end{equation}

\noindent
where
\begin{equation}
\Omega^{\pm}_{rr'}=-{2^{5/2}T\mu^{5/2}\over{ \pi
\omega^2}}\int_{0}^{\infty} dk \int_{0}^{\infty} dx
\int_{0}^{\infty} dy e^{i(px+qy)} \ln
[1+e^{\mu(1-(x+y)^2-k^2\pm\beta y)/T}],
\end{equation}

\noindent
 $p=2\pi \mu r/\omega$, $q=4\pi \mu r'/\omega$, and
$\beta=\omega_c/\omega$. We are interested in  the part of $\Omega$
oscillating with B, which  arises from  terms in Eq.(4) with nonzero
$r,r'$. Replacing $(x+y)$ for $z$ and introducing the
polar-spherical coordinates yield
\begin{equation}
\tilde{\Omega}^{\pm}_{rr'}=-{2^{5/2}T\mu^{5/2}\over{\pi
\omega^2}}\int_{0}^{\infty} dy e^{i(q-p)y}\int_{0}^{\pi/2} d\phi
\int_{y/\cos \phi}^{\infty} d\rho \rho e^{ip\rho\cos\phi}\ln
[1+e^{\mu(1-\rho^2\pm\beta y)/T}].
\end{equation}

\noindent
Replacing $y$ for $y=(\rho^2-\epsilon)/(\pm\beta)$ we obtain
\begin{equation}
\Omega^{+}_{rr'}=-{2^{5/2}T\mu^{5/2}\over{\pi \omega^2}
\beta}\int_{0}^{\infty} d\epsilon e^{i(p-q)\epsilon/\beta}\ln
[1+e^{\mu(1-\epsilon)/T}]\int_{0}^{\pi/2} d\phi
\int_{\sqrt{\epsilon}}^{\gamma(\epsilon,\phi)} d\rho \rho
e^{ip\rho\cos\phi +i(q-p)\rho^2/\beta},
\end{equation}

\noindent
where $\gamma(\epsilon,\phi)=[\epsilon+\beta^2\cos^2(\phi)
/4]^{1/2}+ \beta \cos(\phi)/2$. $\Omega^{-}_{rr'}$ is obtained by
replacing $\beta$ for $-\beta$ in this expression. Here we neglect a
contribution from regions where the total energy $\epsilon$ is
negative. This contribution is an artefact of the approximation (\ref{eqn:calcasymp}), and, in any case, it is exponentially small at temperatures
$T\ll \mu$ in the oscillating part of $\Omega$.

Neglecting terms of the order of $\omega/\mu\ll 1$ the integral over
$\rho$ is approximated as
\begin{equation}
I\equiv\int_{\sqrt{\epsilon}}^{\gamma(\epsilon,\phi)} d\rho \rho
e^{ip\rho\cos\phi +i(q-p)\rho^2/\beta}\approx i\beta
\left[{\sqrt{\epsilon}\,e^{ip\sqrt{\epsilon}\cos\phi+i(q-p)\epsilon/\beta}\over{p\beta\cos\phi+2(q-p)\epsilon}}
-{\gamma(\epsilon,\phi)e^{ip\gamma(\epsilon,\phi)\cos\phi+i(q-p)\gamma^2(\epsilon,\phi)/\beta}\over{p\beta\cos\phi+2(q-p)\gamma(\epsilon,\phi)}}\right]. \label{eqn:appint}
\end{equation}

\noindent
At low temperatures, $T\ll \mu$, the main contribution to the
oscillating part of the thermodynamic potential, $\tilde{\Omega}$,
comes from  energies near the Fermi surface, 
$|1-\epsilon|\ll 1$. Moreover at large $p\gg 1$ only small angles $\phi\ll
1$ contribute to the integral, which allows us to extend the
integration over $\phi$ up to infinity. We can also replace
$\cos(\phi)$ by $1-\phi^2/2$, expand $\gamma(\epsilon,\phi)$ as
\begin{equation}
\gamma(\epsilon,\phi)\approx
\gamma-{{|1-\epsilon|}\over{2(\gamma-\beta/2)}}-{\beta\gamma
\phi^2\over{4[\gamma-\beta/2]}},\label{eqn:gammaapp}
\end{equation}

\noindent
in the exponents (here $\gamma=\sqrt{1+\beta^2/4}+\beta/2$), and
take $\cos\phi=1$, $\gamma(\epsilon,\phi)=\gamma$ in the
pre-exponential terms. Integrating over $\phi$ and $|1-\epsilon|$ with the use
of $\int_0^{\infty} d\phi \exp(ia\phi^2)= (\pi/4|a|)^{1/2} \exp[i\pi
a/(4|a|)]$ and  $\int dx \exp(iax) [1+\exp(x)]^{-1}=-i\pi/\sinh(\pi
a)$ yields
\begin{eqnarray}
\tilde{\Omega}^{+}_{rr'}&=&i{8\pi^{1/2}T\mu^{5/2}\over{\omega^2}}\left[{e^{ip-i\pi
p/4|p|}\over{p|p|^{1/2}\sinh(\pi
pT/2\mu)[p\beta+2(q-p)]}}\right.\nonumber \\
&-&\left.{(\gamma-\beta/2)^{3/2}e^{iq\gamma-i\pi
q/4|q|}\over{q|q|^{1/2}\sinh[\pi qT/2\mu
(\gamma-\beta/2)][p\beta+2(q-p)\gamma]}}\right]
\end{eqnarray}

This approximation fails at the `magic' resonances, where
$p\beta+2(q-p)=0$ or $p\beta+2(q-p)\gamma=0$. In those cases the
integral (\ref{eqn:appint}), is calculated as
\begin{equation}
I\approx (\pi/4|p|)^{1/2}
e^{ip\sqrt{\epsilon}\cos\phi+i(q-p)\epsilon/\beta -i\pi p/4|p|},\label{eqn:res1}
\end{equation}

\noindent
or
\begin{equation}
I \approx \gamma(\pi (\gamma-\beta/2) /4|q|)^{1/2}
e^{ip\gamma(\epsilon,\phi)\cos\phi+i(q-p)\gamma^2(\epsilon,\phi)/\beta-i\pi
q/4|q|},\label{eqn:res2}
\end{equation}

\noindent
respectively.
 Combining equation (\ref{eqn:gammaapp}) and equations (\ref{eqn:res1},\ref{eqn:res2}) we can replace the resonant
 denominators in equation (\ref{eqn:appint}) as
\begin{equation}
{1\over{p\beta+2(q-p)}}\Rightarrow
{1\over{p\beta+2(q-p)+i\beta(4|p|/\pi)^{1/2}\exp(i\pi p/4|p|)}},
\end{equation}

\noindent
and
\begin{equation}
{1\over{p\beta+2(q-p)\gamma}}\Rightarrow
{1\over{p\beta+2(q-p)\gamma-i\beta[4|q|/\pi(\gamma-\beta/2)]^{1/2}\exp(i\pi
q/4|q|)}}.
\end{equation}

\noindent
Substituting this expression into (\ref{eqn:omegapm}) and performing partial
summation yields
\begin{eqnarray}
\tilde{\Omega}&=& \sum_{r=1}^{\infty} \sum_{\pm} A_r^{\pm}\sin(2\pi
r\mu/\omega-\pi/4)+B_{r}^{\pm}\cos(2\pi r\mu/\omega-\pi/4)\cr &+&
C_r^\pm \sin[4\pi r \mu/\tilde{\omega}^\mp-\pi/4]+D_r^\pm \cos[4\pi
r \mu/\tilde{\omega}^\mp-\pi/4]
\end{eqnarray}

\noindent
where \begin{equation} A_r^{\pm}={T(2\omega)^{1/2}\over{2\pi
r^{3/2}\sinh(\pi^2Tr/\omega)}}\Re \cot[\pi r
\omega^{\mp}/\omega\pm\beta (1-i) ( r\omega/\mu )^{1/2}/4],
\end{equation}

\begin{equation}
B_{r}^{\pm}={T(2\omega)^{1/2}\over{2\pi
r^{3/2}\sinh(\pi^2Tr/\omega)}}\Im \cot[\pi r
\omega^{\mp}/\omega\pm\beta (1-i) ( r\omega/\mu )^{1/2}/4],
\end{equation}

\noindent
 and
\begin{equation}
C_r^\pm={T(\tilde{\omega}^{+}+\tilde{\omega}^{-})^{1/2}\over{4\pi
r^{3/2}\sinh[4\pi^2Tr/(\tilde{\omega}^{+}+\tilde{\omega}^{-})]}}\Re
\cot[2\pi r \tilde{\omega}^\pm /(\tilde{\omega}^+ +\tilde{\omega}^-)\pm\beta
(1-i) \omega^2(r/\mu
)^{1/2}/2(\tilde{\omega}^++\tilde{\omega}^-)^{3/2}],
\end{equation}

\begin{equation}
D_r^\pm={T(\tilde{\omega}^{+}+\tilde{\omega}^{-})^{1/2}\over{4\pi
r^{3/2}\sinh[4\pi^2Tr/(\tilde{\omega}^{+}+\tilde{\omega}^{-})]}}\Im
\cot[2\pi r \tilde{\omega}^\pm/(\tilde{\omega}^++\tilde{\omega}^-)\pm\beta (1-i)
\omega^2(r
/\mu)^{1/2}/2(\tilde{\omega}^++\tilde{\omega}^-)^{3/2}].\label{eqn:nlfin}
\end{equation}

\noindent
Here $\omega^\pm=\omega\pm \omega_c/2$, $\tilde{\omega}^\pm=\omega(\sqrt{1+\beta^2/4}\pm \beta/2)$, and
the summation formula, $\sum_{r} (z-r)^{-1}= \pi \cot (\pi z)$ has
been applied. The $+/-$ of $A_r^\pm$ cancel with each other, as do those of $B_r^{\pm}$ -- this allows us to simplify our formula and write: 
\begin{equation}
\tilde{\Omega}= \sum_{r=1}^{\infty} \sum_{\pm} 
C_r^\pm \sin[4\pi r \mu/\tilde{\omega}^\mp-\pi/4]+D_r^\pm \cos[4\pi
r \mu/\tilde{\omega}^\mp-\pi/4]\label{eqn:nlbeg}.
\end{equation}

We can see that, unlike the AK result, we have {\em two} characteristic fequencies $\tilde \omega^{\pm}/2$ which differ from the equivelent frequencies in that result.  In addition, the magic resonances occur at different ratios of $\omega_c/\omega_s$ and are damped by the additional terms in the cotangent function.  As is usual in these calculations, we may restore the effects of spin-splitting by multiplying each term by $\cos(r\pi g\mu_B/e)$.  It should be noted that the effects of spin orbital coupling and other complications may further complicate this expression in real materials.


\section{Numerical results}

Here we present some numerical calculations of the DHVA oscillations and of their Fourier transforms so that the AK and nonlinear results might be compared.  We have set $\mu=2000\omega_s$ and  $\omega_s=1$, measuring $\omega_c$ and all other quantities in units of $\omega_s$.  Fourier transfromations were performed using the NAG DFT routine C06FAF and the Fourier integral calculation techniques described in \cite{NuRes1}.

Calculations at $T=0$ for the parabolic confinement model were performed through taking the limit $T\rightarrow0$ in (\ref{eqn:thrmpot}) and then integrating over $k$.  We thus obtained the following expression for the thermodynamic potential, 
\begin{equation}
\Omega=-\frac{4\sqrt{2}}{3\pi}\sum_{n,\, m} \Re(\mu-E_{nm})^{\frac{3}{2}},
\end{equation}

\noindent where $E_{nm}$ is energy for a given $n,\,m$,
The non-oscillatory portion was then subtracted out, leaving only the oscillatory portion of the function behind.  

This could not be done in the case of the non-linear model, however, due to the issue remarked on in Section \ref{sec:nlDHVA} regarding the unphysical negative energy levels.  It appears that as yet there is no straightforward way of calculating the thermodynamic quantities in this fashion, as it does not appear as easy to remove the contributions from negative energy values in this method as it was in the aforementioned analytic (`Poisson summation') treatment of Section  \ref{sec:nlDHVA}.

For $T>0$, the calculations were done using the formulae of \cite{Alexandrov1} for the AK results, and (\ref{eqn:nlbeg})-(\ref{eqn:nlfin}) in this paper were used for the non-linear results.

\begin{figure}[htb]
\vspace{0.5cm}
\begin{center}
\epsfig{file=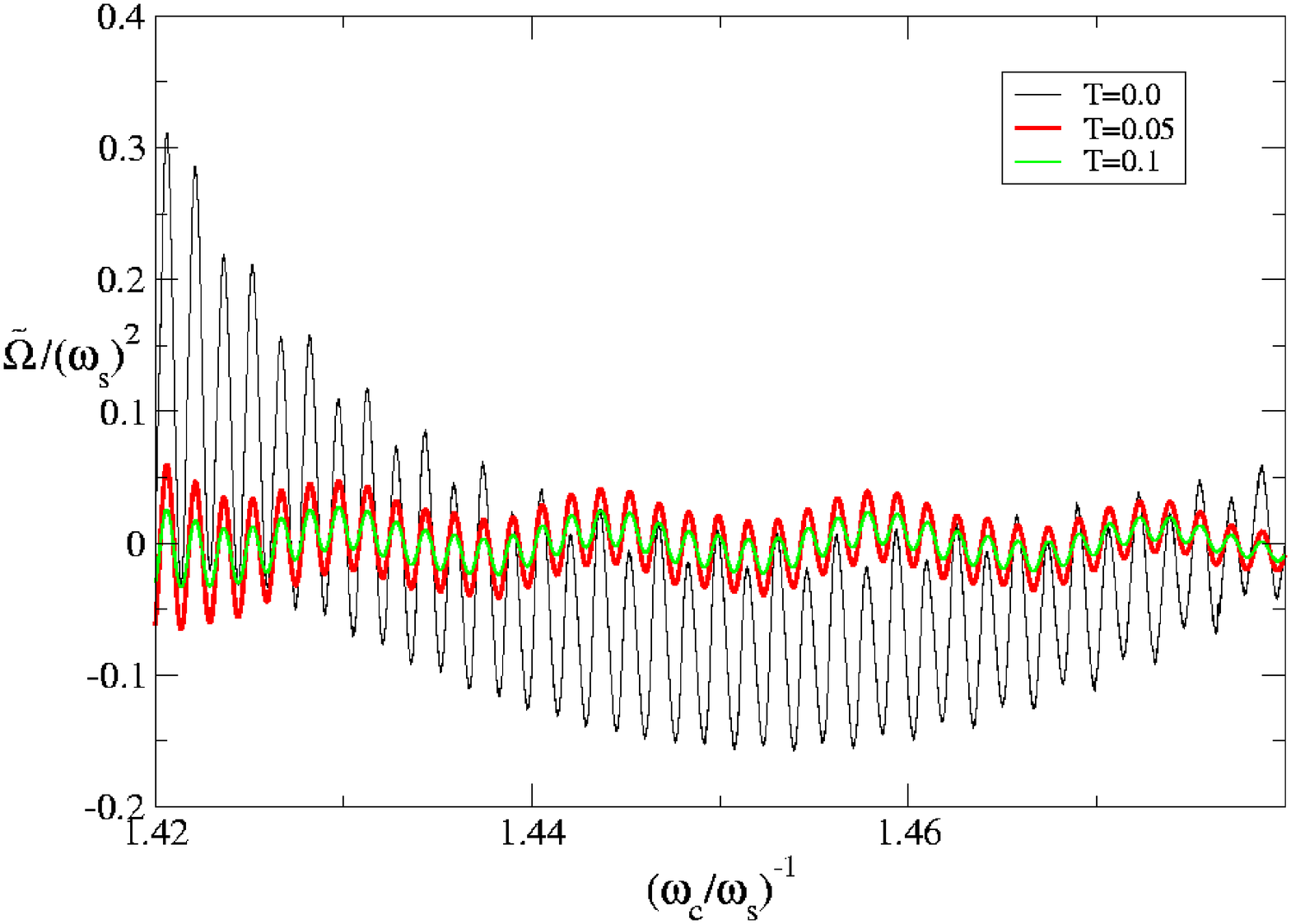,height=5.4cm}
\hspace{.5cm}
\epsfig{file=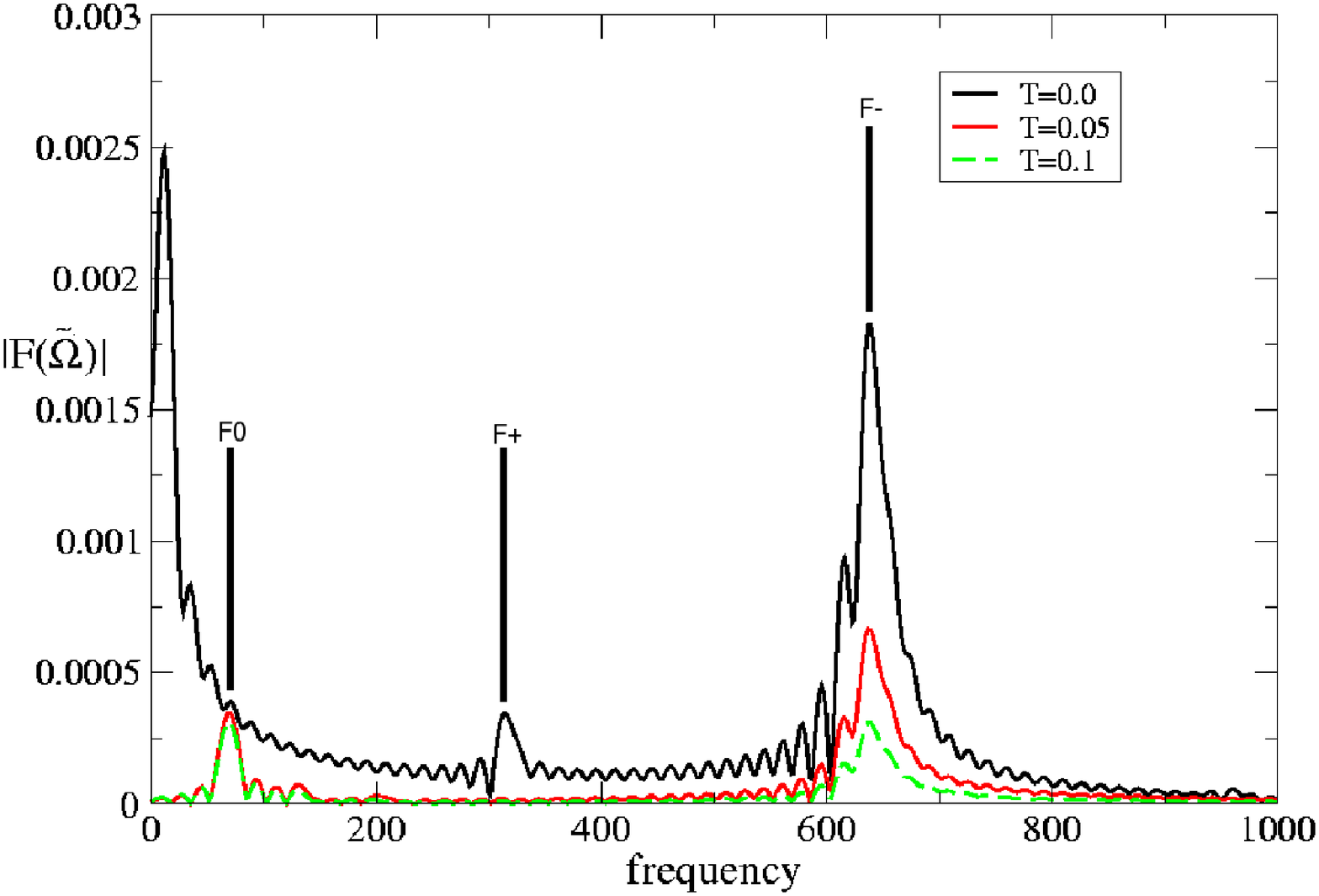,height=5.4cm}
\end{center}
\vspace{.3cm}
\begin{center}
\epsfig{file=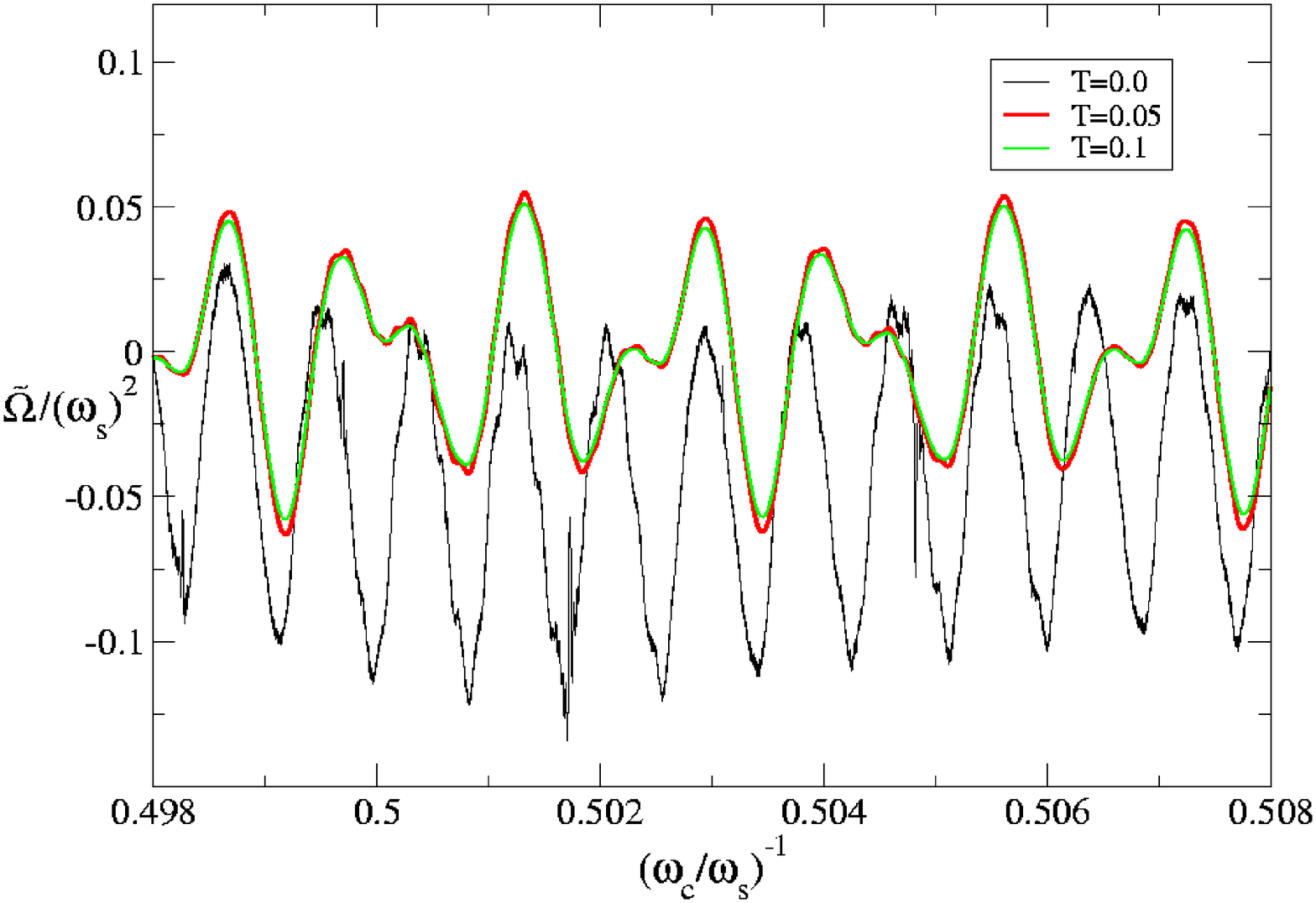, height=5.3cm}
\hspace{.4cm}
\epsfig{file=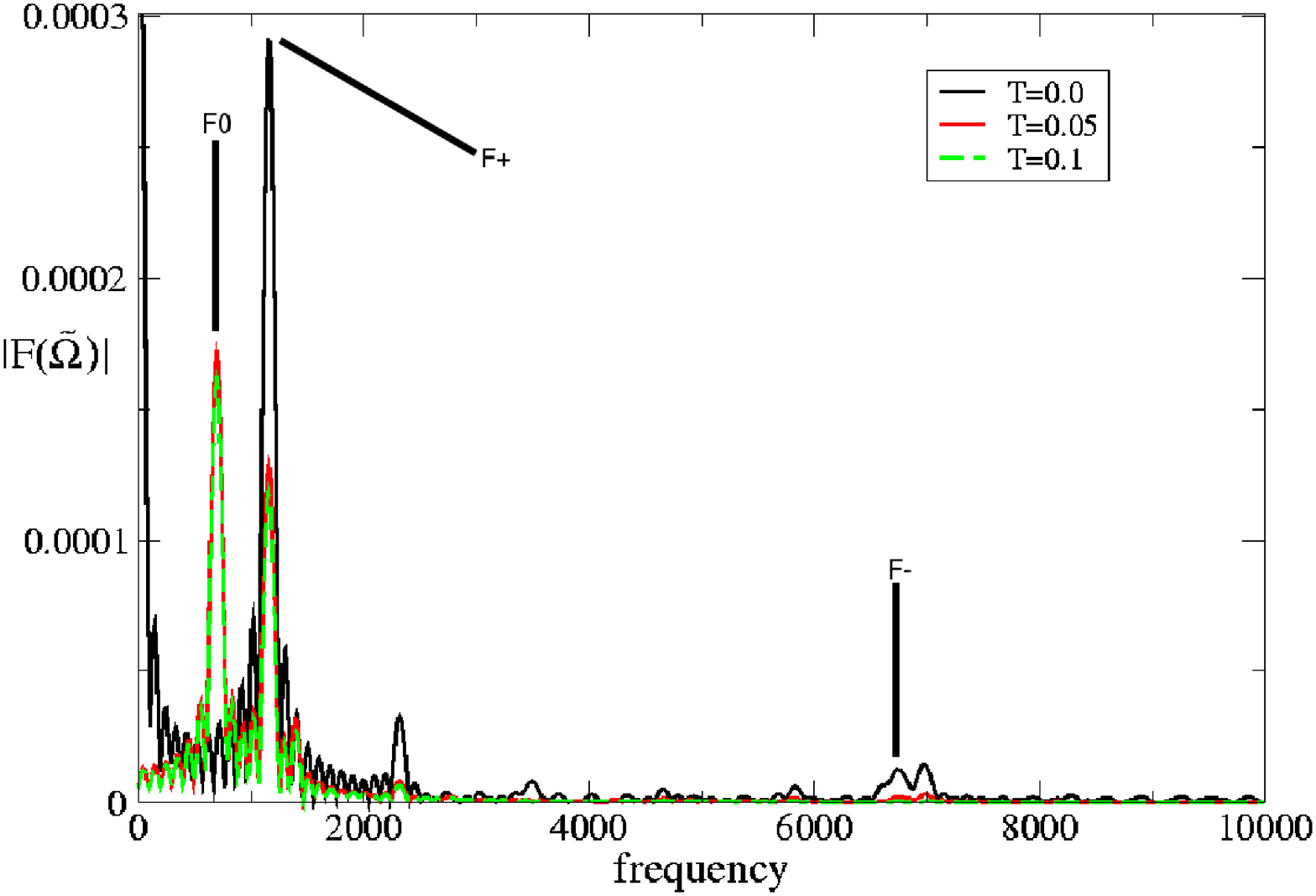,height=5.3cm}
\end{center}
\vspace{-.5cm}
\caption{\small Graphs of $\tilde\Omega$ in the parabolic confinement model opposite their Fourier transforms.  Note that the remnant of the non-periodic component of the $T=0$ data is most noticeable in the topmost graph, but also that the $F+$ resonance is most obviously visible in its Fourier transform.}
\label{fig:FDomega}
\end{figure}

Figure \ref{fig:FDomega} displays the behavior of the thermodynamic potential  with respect to changes in $(\omega_c)^{-1}$, alongside their Fourier transforms. These last clearly show the three peaks corresponding to each of the characteristic frequencies given in \cite{Alexandrov1} (here, $F_0$ is equivalent to the $F$ of that reference).  In the case of the $T=0$ plots it seems that some residual non-oscillatory components of the untransformed functions remain at low frequencies, and these obscure the low frequency content of the Fourier transform. 

\begin{figure}[htb]
\vspace{0.5cm}
\begin{center}
\epsfig{file=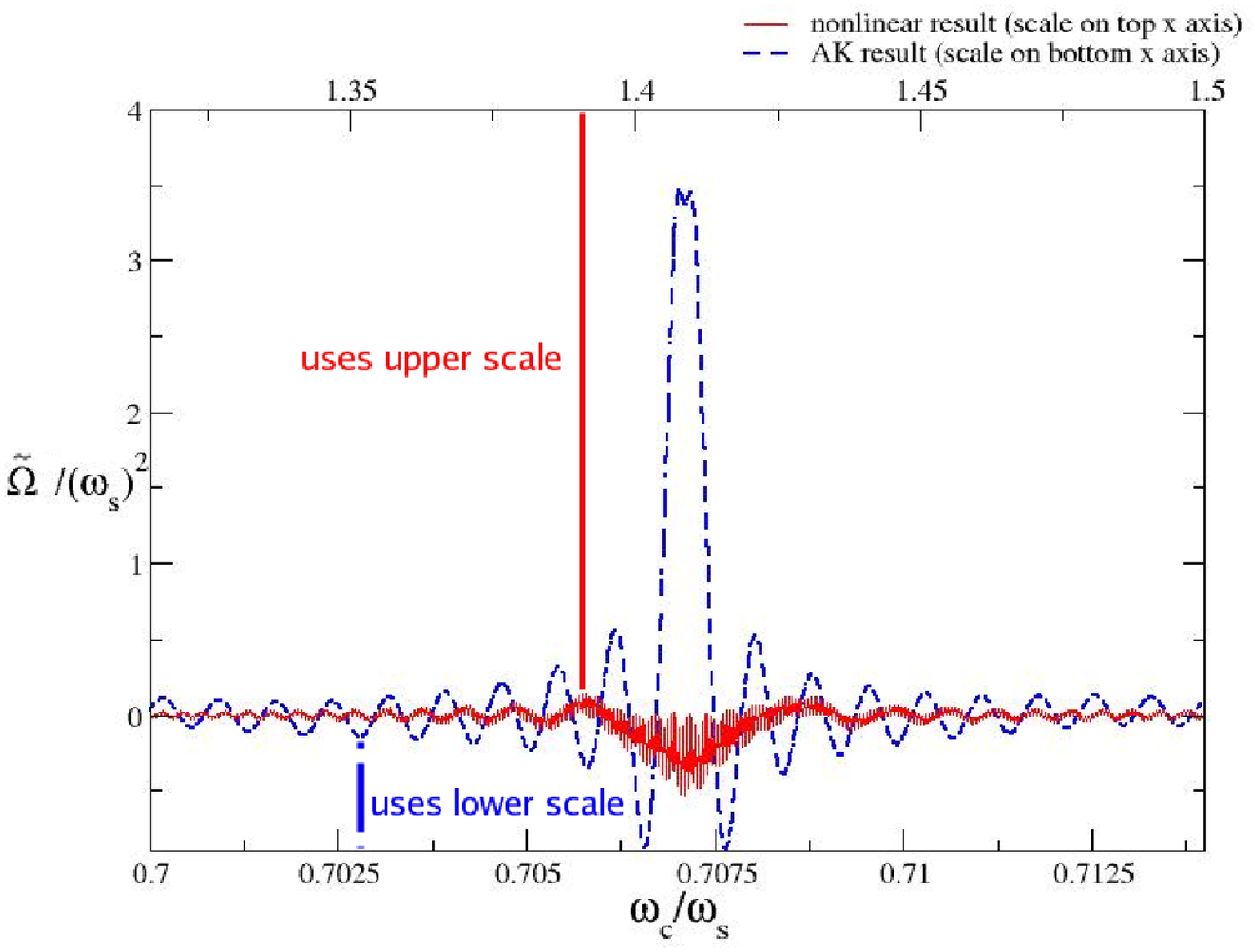, height=7cm}
\end{center}
\vspace{-.5cm}
\caption{\small Graph comparing the first magic resonances of the parabolic confinement model with the non-linear model.  The temperature here is set at $T=0.025$.}
\label{fig:nlvsFD}
\end{figure}

In Figure \ref{fig:nlvsFD} we present the first magic resonance in the thermodynamic potential of the non-linear model together with that of the parabolic confinement model for the purposes of comparison.  We can see that the resonance in the non-linear case has not only changed position, but has also been damped by the additional terms in the cotangent functions of the Fourier coefficients.  

\begin{figure}[htb]
\vspace{0.5cm}
\begin{center}
\epsfig{file=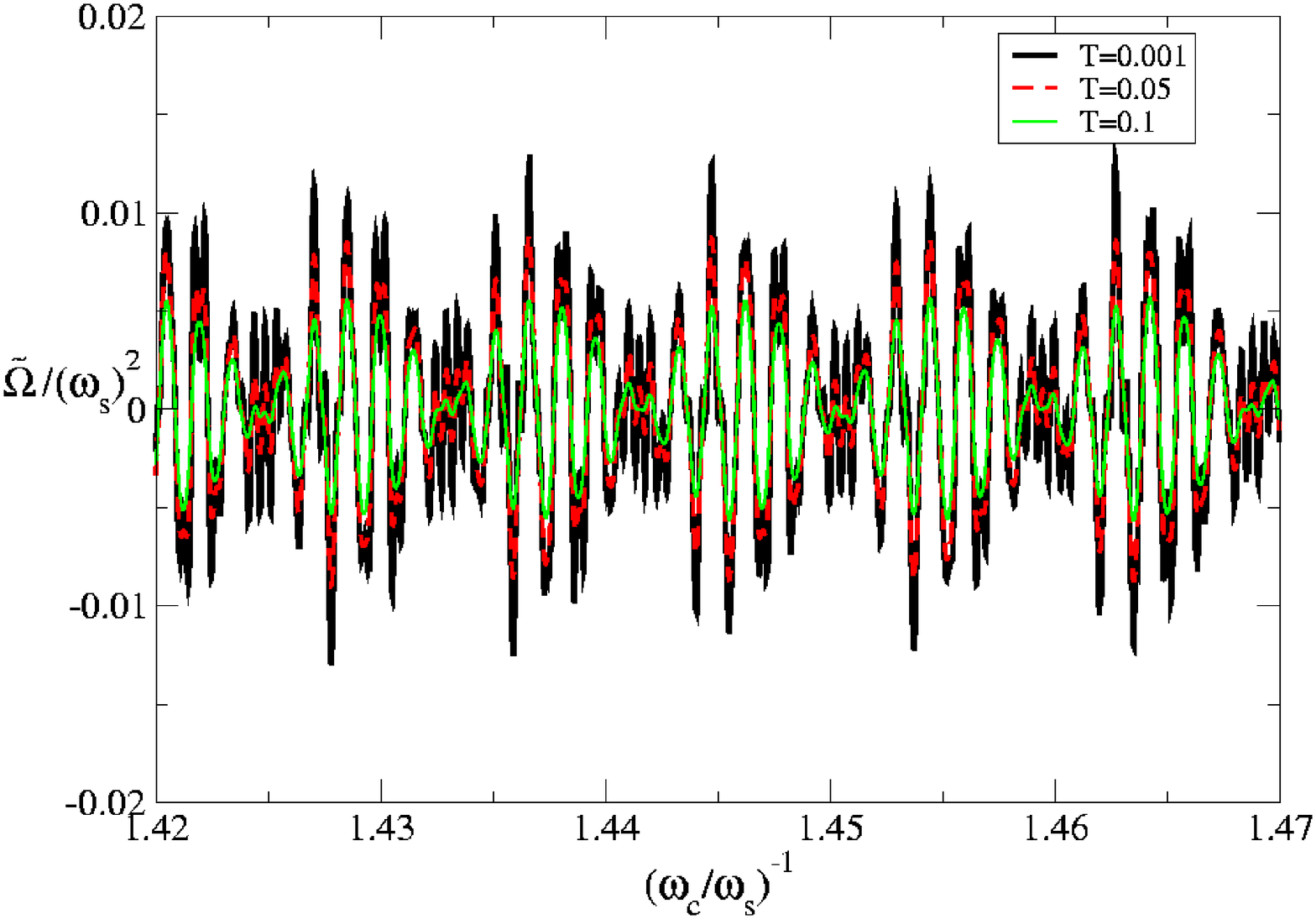, height=5.4cm}
\hspace{.0cm}
\epsfig{file=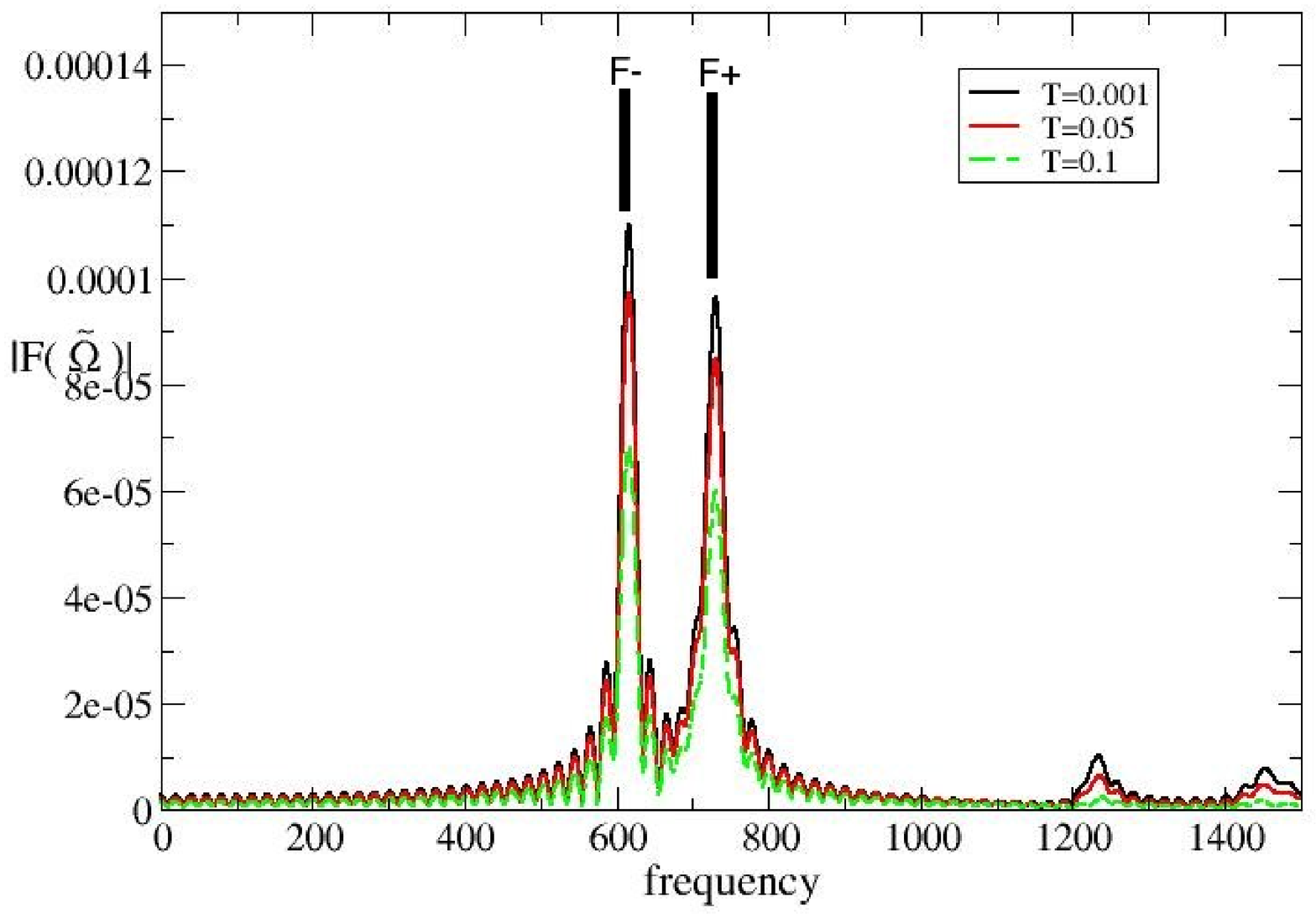,height=6.4cm}
\end{center}
\vspace{.3cm}
\begin{center}
\epsfig{file=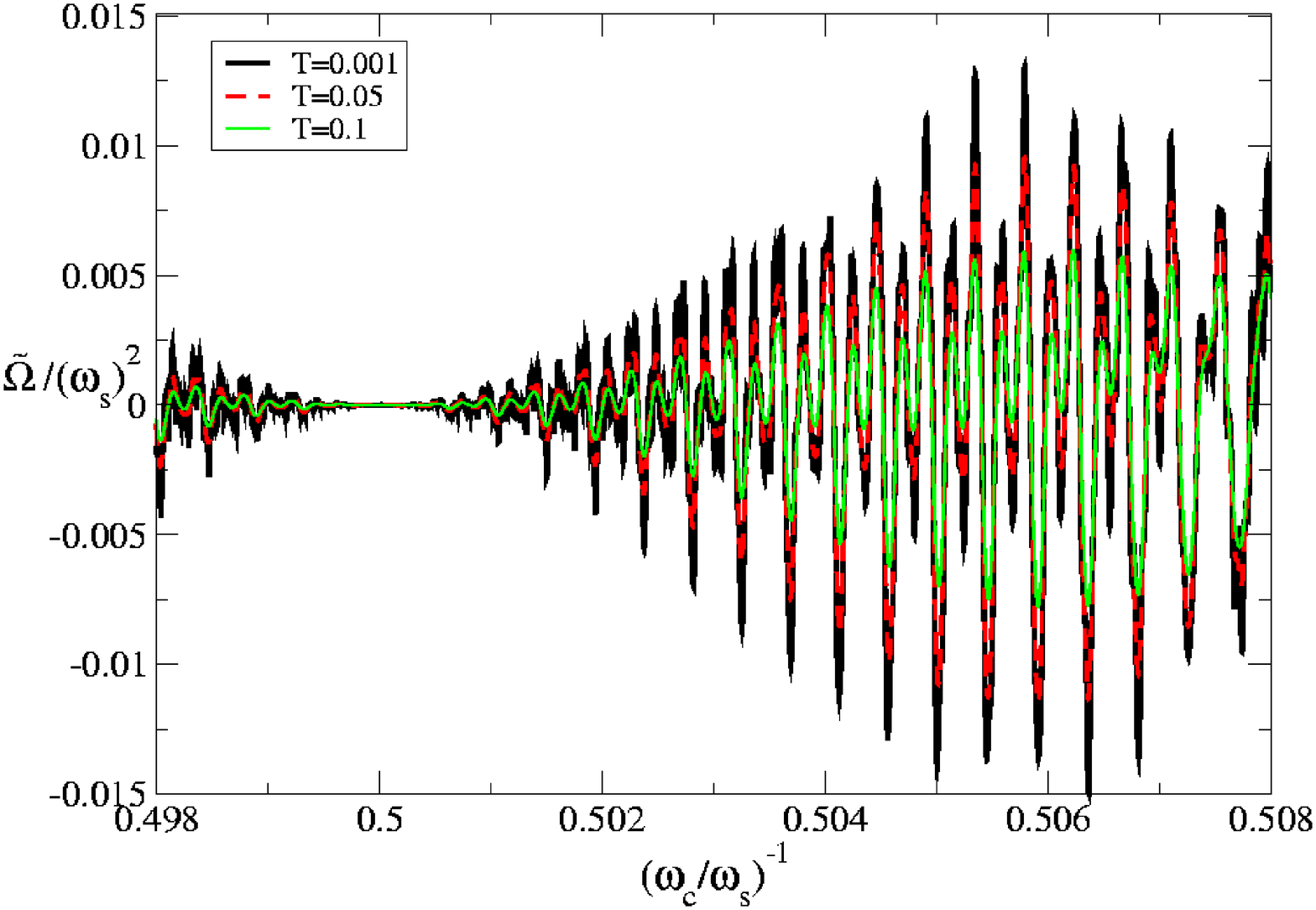, height=5.3cm}
\hspace{.0cm}
\epsfig{file=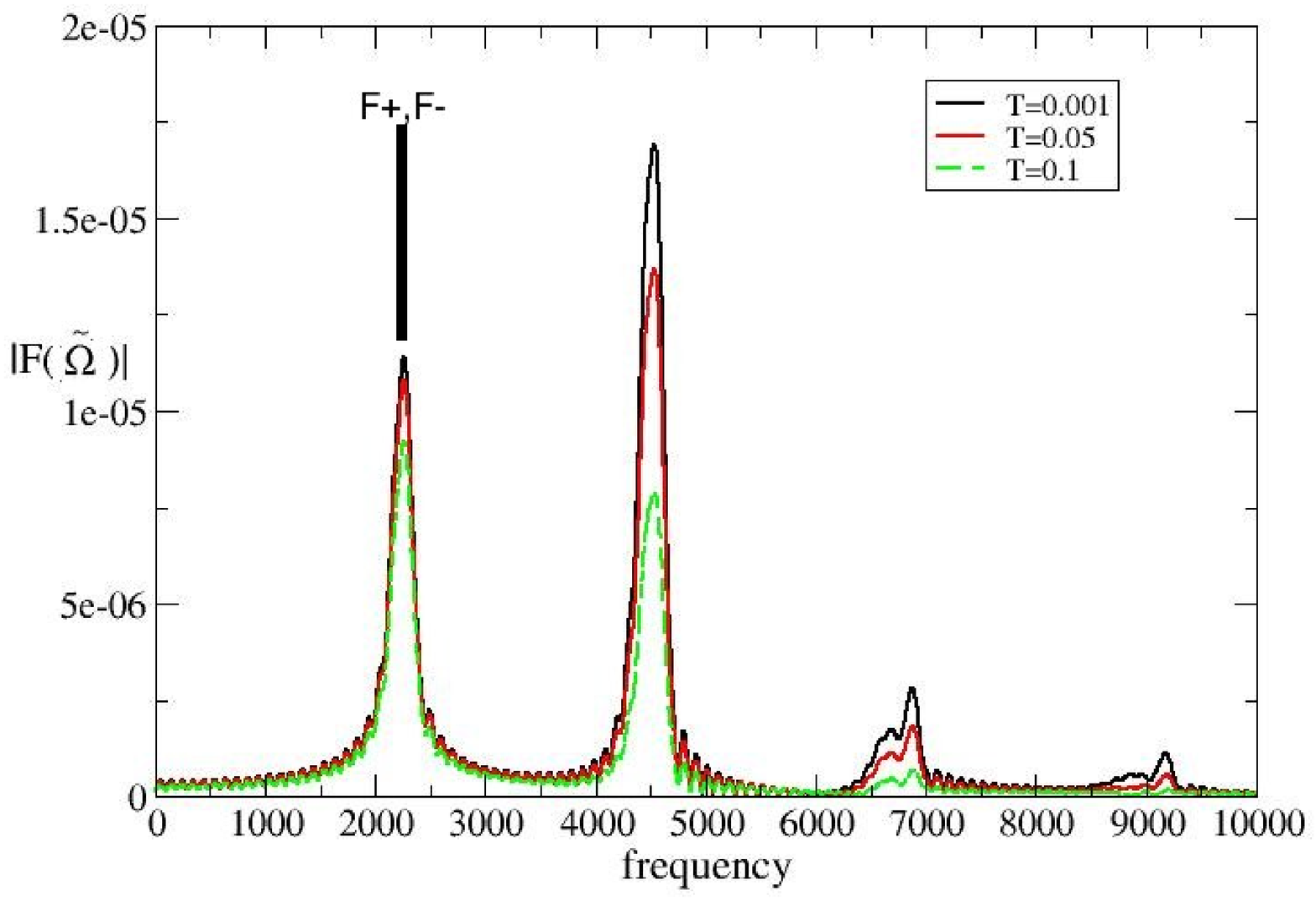,height=6.4cm}
\end{center}
\vspace{-.5cm}
\caption{\small Graphs of $\tilde \Omega$ in the non-linear model opposite their Fourier transforms.  Note that the $F+$ and $F-$ peaks in the lower Fourier transform are so close as to be indistiguishable; this may be why the second harmonic is larger than expected.}
\label{fig:nlomega}
\end{figure}

Figure \ref{fig:nlomega} shows the thermodynamic potential and its Fourier transform in the same ranges as previously used for the parabolic confinement model calculations.  Following \cite{Alexandrov1}, we may estimate the characteristic frequencies as being:

\begin{equation}
F^\pm=\frac{2F |\sqrt{2}\pm\sqrt{1+\gamma'}|}{\sqrt{1+\gamma'}(\sqrt{1+\gamma'}\pm(1/\sqrt{2}))^2},\end{equation}                   

\noindent where $F= S_F/(2\pi e)$, $\gamma=4\omega_s^2/\omega_c^2=4\pi^2 S_F/(e^2SB^2)$, $\gamma'=\gamma/2$, $S=\pi R^2$ is the cross-sectional area of the wire, and $S_F=\pi (m^*)^2v_F^2$ is the cross-sectional area of the Fermi-surface.  (We remind the reader that in order to interpret the frequencies given in the figures using these formulae, we should take $e=m^*=\omega_s=1$).

We can see the two characteristic frequencies clearly in the Fourier transform,  and that the oscillatory behavior of the function is considerably different from that seen in the parabolic confinement case.

\section{How good are the approximations?} \label{sec:errors}

In order to understand how our non-linear model differs from the exact behavior of the system, we must compare their spectra with that generated from the exact zeroes of the confluent hypergeometric function $M(a,b,\xi)$.  In order to obtain the latter, we numerically generated $M(a,b,\xi)$ with the CONHYP function \cite{Nardin1,Nardin2}, and located the roots of the function as the parameter $-a$ was increased  from zero using the Van Wijngaarden-Dekker-Brent algorithm \cite{NumRes}.

Unfortunately, the processing ability of the computer limits the values of the variable $\xi$ for which results can be obtained; this means that it is difficult to obtain zeroes for large values of both $\mu/\omega_s$ and $\omega_c/\omega_s$, since this entails a large value of $\xi$ that will cause the series defining $M(a,b,\xi)$ to be very slowly converging.  However, one can gain an idea of the behavior of the exact system for less extreme values of the variable, and it is to such an example that we now turn.

We set $\mu=100\omega_s$, $\omega_s=1$ and examine values of $0\le{\omega_c}/{\omega_s}\le1$.  For each of these, we determine the largest value of $n$ in the non-linear approximation for which $E\le\mu$ (assuming $k$ is zero) and calculate the relative error $\Delta=(E-E_{exact})/E$, where $E_{exact}$ is the exact energy calculated from the $n$-th zero of $M(a,b,\xi)$.  This enables us to gain some idea of the size of the error at the Fermi surface described by the non-linear model.

To obtain the exact energy in the zero-field limit, we note that the wavefunction in that case takes the form of a Bessel function, and so the exact zeroes may be obtained from the function $J_{|m|}(2\sqrt{E_{exact} w})$, where $w=(\frac{\pi}{2\omega_s})^2\mu$.
\nopagebreak

\begin{figure}
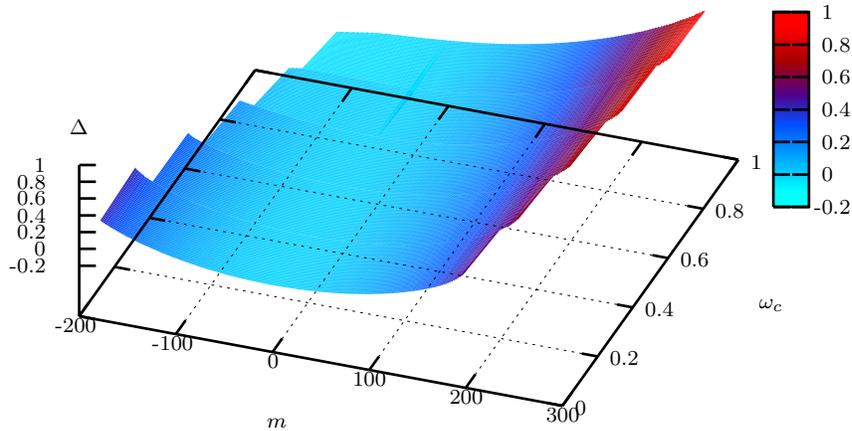

\vspace{0.5cm}
\begin{center}
\include{relerr1}
\end{center}
\vspace{-.5cm}
\caption{\small The relative error $\Delta$ along the Fermi surface defined by the non-linear model, as a function of $m$ and $\omega_c$.}
\label{fig:fermerr}
\end{figure}
\nopagebreak

\begin{figure}[htb]
\vspace{0.5cm}
\begin{center}
\epsfig{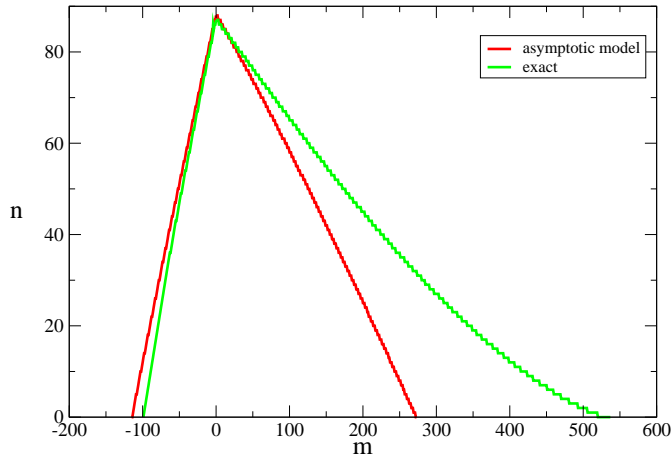}
\end{center}
\vspace{-.5cm}
\caption{\small Graph showing the quantum number $n$ against $m$ at the Fermi surface for $\mu=100\omega_s$ and $\omega_c/\omega_s=1.0$.}
\label{fig:fermserf}
\end{figure}

We plot $\Delta$  in Figure \ref{fig:fermerr}.  It is apparent that as $\omega_c$ increases, the shape of the approximate Fermi surface alters such that the maximum possible positive value of $m$ increases, and the maximum negative value of $m$ decreases.  This coincides with a decrease in the maximum value of $|\Delta|$ for negative $m$, and an increase for positive $m$ -- in fact the latter error is so large for large $m$ at $\omega_c=1$ that it can scarcely be said that it accurately models the exact values at all in that region.  In addition, it is worth noting that the non-linear model {\em underestimates} the deformation of the Fermi surface when $\omega_c$ is large (for example, see Figure \ref{fig:fermserf}).  Why is this?

In the case of the asymptotic derivation of the model, the asymptotic from which we obtained it makes two assumptions:  that $-a$ is greater than $\xi$, so that $M(a,b,\xi)$ resembles a Bessel function, and that $-a$ is greater than $b$, so that the Bessel function resembles a cosine function.  Obviously, this condition is violated in the zero-field limit at large $m$, and so we would expect it to be violated for large $m$ at finite values of the field, even though it appears that the replacement of $\omega_s$ with $\omega$ in Section \ref{sec:connect} counteracts the violation of the first condition for large $\xi$ if $b$ is small. This is what occurs in the case of positive $m$.  As for the the portions of the spectrum with negative $m$, since these have larger energies than their positive counterparts and since this disparity increases as we increase $\omega_c$, one should not be surprised to find that, as, $\omega_c$ increases, the Fermi energy is exceeded at smaller and smaller values of negative $m$, which are in turn more closely matched by the asymptotic.  And so in this case the maximum error decreases.

Semiclassically, one should note that large values of $m$ describe electron orbits that approach the edge of  the wire quite closely.  Since that is a turning point due to the imposition of the `hard' boundary conditions, we cannot expect the semiclassical approximation to hold in that region.   The analysis of the difference between the behavior at positive and negative $m$ given above also holds here.

Perhaps we should consider the boundary conditions of our problem.  The parabolic confinement model was originally derived with `soft' boundary conditions -- a parabolic confining potential -- and as we have shown, this approximates the non-linear model, derived with `hard' boundary conditions near the Fermi surface.  However, this latter model does not capture the large $m$ behavior of the system very well, in part, one suspects, because of the `hardness' of the boundary conditions.  It should be noted then, that neither of these extremes is likely to be physical -- the boundary conditions for a real nanowire will most probably be intermediate between them, and so it is possible that the experimental behavior of the system might also be intermediate between the behaviors we have discussed.  This, of course, is a matter that requires empirical determination.


\section{Conclusions}

We have calculated the oscillations in the thermodynamic potential due to the De Haas-Van Alphen effect in a clean, metallic nanowire with a simple Fermi surface for  a nonlinear energy spectrum derived from an asymptotic approximation to the exact solution of the Schr\"odinger equation, and compared it with the AK results derived using the parabolic confinement model which approximates it  near to the Fermi energy.  In both models, one can observe `magic resonances' where the amplitude of the oscillations in the thermodynamic potential is enhanced at particular ratios of $\omega_s$ to $\omega_c$, although  their locations differ in each model, and are somewhat damped in the non-linear case.  However, the nonlinear result lacks one of the frequencies of oscillation observed in the AK result.  In addition, it seems that the infinite well boundary conditions  cause some problems at large values of $m$ due to the semiclassical nature of our approximations; however, this is unlikely to be fatal to our predictions of magic resonances since the real system likely has boundary conditions intermediate between the two cases which we have considered.  

It seems clear that the interplay between the size quantization and the magnetic quantization is non-trivial in the extreme, and exhibits a certain degree of model dependence.  Further experimental and theoretical work is needed both to clarify remaining ambiguities, and to apply this general theory to the more complex band structures of realistic nanowires.

\section{Acknowledgments}

The authors would like to thank the EPSRC for funding this research (grant No. EP/D035589), and David Khmelnitskii for helpful comments.



\end{document}